# Factors determining surface oxygen vacancy formation energy in ternary spinel structure oxides with zinc

Yoyo Hinuma[1,2*], Shinya Mine[3], Takashi Toyao[3,4], Takashi Kamachi[4,5], and Ken-ichi Shimizu[3,4]

[1] Department of Energy and Environment, National Institute of Advanced Industrial Science and Technology (AIST), 1-8-31, Midorigaoka, Ikeda 563-8577, Japan

[2] Center for Frontier Science, Chiba University, 1-33 Yayoicho, Inage, Chiba 263-8522, Japan

[3] Institute for Catalysis, Hokkaido University, N-21, W-10, Kita, Sapporo, Hokkaido 001-0021, Japan

[4] Elements Strategy Initiative for Catalysts and Batteries, Kyoto University, Katsura, Nishigyo, Kyoto 615-8520, Japan

[5] Department of Life, Environment and Applied Chemistry, Fukuoka Institute of Technology, 3-30-1Wajiro-Higashi, Higashi-ku, Fukuoka 811-0295, Japan

* y.hinuma@aist.go.jp

## Abstract

Spinel oxides are an important class of materials for heterogeneous catalysis including photocatalysis and electrocatalysis. The surface O vacancy formation energy ($E_{\text{Ovac}}$) is a critical quantity on catalyst performance because the surface of metal oxide catalysts often acts as reaction sites, for example, in the Mars-van Krevelen mechanism. However, experimental evaluation of $E_{\text{Ovac}}$ is very challenging. We obtained the $E_{\text{Ovac}}$ for (100), (110), and (111) surfaces of normal zinc-based spinel oxides $ZnAl_2O_4$, $ZnGa_2O_4$, $ZnIn_2O_4$, $ZnV_2O_4$, $ZnCr_2O_4$, $ZnMn_2O_4$, $ZnFe_2O_4$, and $ZnCo_2O_4$. The most stable surface is (100) for all compounds. The smallest $E_{\text{Ovac}}$ for a surface is the largest in the (100) surface except for $ZnCo_2O_4$. For (100) and (110) surfaces, there is a good correlation, over all spinels, between the smallest $E_{\text{Ovac}}$ for the surface and bulk formation energy, while the ionization potential correlates well in (111) surfaces. Machine learning over $E_{\text{Ovac}}$ of all surface sites in all orientations and all compounds to find the important factors, or descriptors, that decide the $E_{\text{Ovac}}$ revealed that bulk and surface-dependent descriptors are the most important, namely the bulk formation energy, a Boolean descriptor on whether the surface is (111), and the ionization potential, followed by geometrical descriptors that are different in each O site.

## 1. Introduction

Defects can significantly influence the properties of metal oxides, where the O vacancy is the most representative defect.[1, 2] O vacancies, when intentionally or unintentionally introduced into the metal oxide structure, could strongly affect the electrical, optical, magnetic, mechanical, and catalytic properties.[3, 4] Surface point defects substantially affect heterogeneous catalysis because O vacancies on the surface of metal oxide catalysts often act as reaction sites.[5, 6] As a consequence, the formation energy of an O vacancy at the surface, which is denoted as $E_{Ovac}$ in this paper, is often used as a descriptor of the catalytic activity of metal oxides.[7-10]

However, experimental investigations of O vacancies remain formidable tasks despite the obvious importance of research on O vacancies in the field of catalysis.[11, 12] Highly sophisticated techniques are necessary for determination of $E_{Ovac}$ and the evaluation of $E_{Ovac}$ is not always possible.[13] On the other hand, several theoretical studies on the formation of O vacancies on metal oxides have recently emerged.[14-20] The number of investigated surfaces still remains limited although a number of contributions were made to obtain $E_{Ovac}$ values of metal oxide surfaces. Therefore, studies that comprehensively reveal the physical principles determining $E_{Ovac}$ are highly desirable. In this sense, we have recently reported $E_{Ovac}$ of various insulating and semiconducting simple binary oxide surfaces using DFT calculations at the same computational level and comparable structure models.[21]

The spinel structure with the composition $AB_2X_4$ constitutes one of the most important classes of crystalline compounds in catalysis.[22-26] In the normal spinel structure, A and B cations occupy two different sites in their structure, namely octahedral and tetrahedral sites, without mixing. In contrast, the octahedral sites are occupied by both A and B cations in the inverse spinel structure. The spinel oxides have attracted much attention in the field of heterogeneous catalysis including photocatalysis and electrocatalysis thanks to their structure diversities where their characteristics can be tailored by choosing appropriate A and B cations.[27-29] For example, Zn-based spinel oxides are used as catalysts in various $CO_2$ hydrogenation reactions [30-33] and here O vacancies play significant roles.

In this paper, the $E_{Ovac}$ for (100), (110), and (111) surfaces of eight normal zinc-based spinel oxides without mixing of cations between A and B sites, namely $ZnM_2O_4$ where M is the B-site cation and is one of Al, Ga, In, V, Cr, Mn, Fe, or Co, were evaluated and the existence or non-existence of correlations with other physical quantities were investigated.

## 2. Computational methods

First-principles calculations were conducted using the projector augmented-wave method[34] as implemented in the VASP code[35, 36]. The PBEsol functional[37] was used among the generalized gradient approximations (GGAs) because it reasonably reproduces energetics and crystal structures in oxide systems[38], for instance, compared to the standard Perdew, Burke and Ernzerhof (PBE) functional[39]. This work focuses on $ZnM_2O_4$ where M is one of Al, Ga, In, V, Cr, Mn, Fe, or Co. These compounds are experimentally known to take the normal spinel structure without exchange of species between the tetrahedral Zn and octahedral M sites [40, 41]. Choosing normal spinel structures only avoids the problem of dealing with partial occupancies on the octahedral sites. The effective $U$ value of $U$-$J$ for the valence $3d$ states was set at 5 eV for Zn and 3 eV for V, Cr, Mn, Fe, and Co. These values are the same as in Ref. [42]. The spin states of M with a formal charge of 3+ are high spin $d^2$ in $V^{3+}$, high spin $d^3$ in $Cr^{3+}$, high spin $d^4$ in $Mn^{3+}$, high spin $d^5$ in $Fe^{3+}$, and low spin $d^6$ except some undercoordinated ions at the surface that can have non-zero spin in $Co^{3+}$. Among these, species subject to Jahn-Teller distortion are $V^{3+}$ and $Mn^{3+}$. The effect of adding the Hubbard $U$ was considered based on Dudarev's formalism.[43] As the magnetic ordering and the energy difference between the different magnetic solutions are highly sensitive to the functional chosen, PBEsol+$U$ may not be adequate and hybrid functionals might be necessary. However, we stick to practical PBEsol+$U$ calculations in this study.

The highest symmetry space group type of normal spinel is $Fd\bar{3}m$ (number 227), and all surfaces of normal spinel in this space group are polar (type 3) according to Tasker's definition[44], therefore making a nonpolar slab where both surfaces are identical may appear impossible. However, it is actually possible because the surfaces are nonpolar type C in the definition by Hinuma *et al.* [45]. Although it is impossible to obtain a nonpolar and stoichiometric slab by simply cleaving bulk, one can be obtained by removing half of the atoms in the outermost surfaces, for instance in a stripe pattern[46].

Defect formation (O desorption) were performed on both sides of a slab such that the slab is always nonpolar. Internal coordinates and lattice parameters were relaxed in bulk calculations, and all internal coordinates were allowed to relax while lattice parameters were fixed in slab calculations.

The surface energy $E_{\text{surf}}$ is defined as $E_{\text{surf}} = (E_{\text{slab}} - E_{\text{bulk}})/2A$, where $E_{\text{slab}}$, $E_{\text{bulk}}$, and $A$ are the energy of the slab without defects, the energy of the constituents of the slab when in a perfect bulk, and the area of one side of the slab respectively. The O vacancy formation energy is defined as $E_{\text{Ovac}} = \left(E_{\text{removed}} - E_{\text{slab}} + 2\mu_O\right)/2$, where $E_{\text{removed}}$ and $\mu_O$ are the energy of the slab when two O atoms are removed (one O from each surface) and the chemical potential of the O that is removed, which is referenced to O$_2$ gas at 0 K, respectively. $E_{\text{surf}}$ was calculated for both "thin" and "thick" slabs (details given in Supplementary Table SI-1), and the values for these slabs were linearly extrapolated to obtain the "fitted" $E_{\text{surf}}$ at the zero slab thickness limit (see Ref. [21] for details).

The primitive cell contains 14 atoms, where there are two, four, and eight symmetrically equivalent Zn, M, and O atoms, respectively. However, the energies of ZnM$_2$O$_4$ (M = V, Cr, Mn, or Fe) are lower, compared to a ferromagnetic spin ordering, in an antiferromagnetic spin ordering when the four M atoms in the primitive cells are divided into two spin up atoms and two spin down atoms. The energy difference is 72, 16, 7, and 1 meV/formula unit for M = V, Cr, Mn, and Fe, respectively. Introducing this antiferromagnetic spin ordering by considering the up and down spin M as two distinct species lowers the space group type to *I*4$_1$/*amd* (number 141) where the spin alternates along the *c* axis. ZnV$_2$O$_4$ is dynamically stable at this space group type. However, ZnMn$_2$O$_4$ requires further lattice distortion and is dynamically stable in the space group type *I*4$_1$/*a* (number 88). Bulk properties of the considered structures are summarized in Table 1 together with experimentally reported band gaps (BGs). Calculations typically underestimated the BGs, although the calculated BG overestimated the experimental value in ZnAl$_2$O$_4$.

A different magnetic ordering was suggested in ZnV$_2$O$_4$ by Reehis et al.[47]. This magnetic ordering requires 28 atoms in the primitive cell. The space group type corresponding to this magnetic ordering is *F*222 (number 22), and its formation energy is almost the same with difference of at most 16 meV/atom from the 14-atom primitive cell. Slabs for systems with antiferromagnetic trivalent atoms were obtained assuming that up-spin and down-spin cations are different species. The 14-atom primitive cell was adopted because a reasonable termination for the (110) and (111) surfaces cannot be obtained with the proposed algorithm. The magnetic ordering of slab models based on the 28-atom primitive cell is incommensurate with the periodicity of slab models based on the 14-atom primitive cell, therefore relaxation of the magnetic ordering does not happen.

**Table 1.** Bulk properties of ZnM$_2$O$_4$. $v$, $E_{\text{bulk}}$, and band gap (BG) are the volume per atom, formation energy per atom, and minimum band gap, respectively. Experimentally reported BGs are also shown.

| Compound | Lattice parameters | | | $v$ ($\text{Å}^3$/atom) | $E_{\text{bulk}}$ (eV/atom) | BG (eV) | Experimental BG (eV) |
|---|---|---|---|---|---|---|---|
| | $a$ (Å) | $c$ (Å) | $c/a$ | | | | |
| ZnAl$_2$O$_4$ | 8.049 | | | 9.310 | -2.81 | 4.45 | 3.8-3.9 [48] |
| | | | | | | | 4.1-4.3 [48] |
| ZnGa$_2$O$_4$ | 8.329 | | | 10.318 | -2.00 | 2.76 | Direct 4.59 ± 0.03, indirect 4.33 [49] |
| ZnIn$_2$O$_4$ | 8.937 | | | 12.745 | -1.67 | 1.36 | |
| ZnV$_2$O$_4$ | 5.931 | 8.444 | 1.424 | 10.608 | -2.51 | 0.55 | 2.8 [50] |
| ZnCr$_2$O$_4$ | 5.886 | 8.257 | 1.403 | 10.218 | -2.20 | 2.42 | 3.35 [51] |
| ZnMn$_2$O$_4$ | 6.049 | 8.016 | 1.325 | 10.476 | -2.00 | 0.81 | 1.51 [52] 1.58 [53] 2.1 [54] |
| ZnFe$_2$O$_4$ | 5.900 | 8.382 | 1.421 | 10.420 | -1.71 | 1.13 | 2.15 [55] 2.316 [56] |
| ZnCo$_2$O$_4$ | 7.986 | | | 9.080 | -1.56 | 1.89 | 1.82 [57] |

## 3. Results and discussion
### 3.1 Convergence of surface properties

When discussing the ($hkl$) surface, a supercell with basis vectors $(\mathbf{a}', \mathbf{b}', \mathbf{c}')$ is considered where $\mathbf{a}'$ and $\mathbf{b}'$ are "in-plane vectors". Taking $(\mathbf{a}, \mathbf{b}, \mathbf{c})$ as the basis vectors of the conventional cell, an in-plane vector $h'\mathbf{a} + k'\mathbf{b} + l'\mathbf{c}$ satisfies $(h', k', l') \cdot (h, k, l)^{\text{T}} = 0$ The thickness of the supercell is defined as $h_{\text{cell}} = \dfrac{(\mathbf{a}' \times \mathbf{b}') \cdot \mathbf{c}'}{|\mathbf{a}' \times \mathbf{b}'|}$ (length of the out-of-plane basis vector projected in the direction normal to the surface) and the slab and vacuum thicknesses are defined as $h_{\text{slab}} = r h_{\text{cell}}$ and $h_{\text{vac}} = (1-r) h_{\text{cell}}$, respectively, where $r$ is the ratio of atoms remaining after atom removal from a completely filled supercell. Supplementary Table 1 gives information on the geometry of employed slabs. We investigated the most stable termination among the (100), (110), and (111) orientations in the cubic setting (orientations are referenced to the cubic setting throughout this paper). Various surface terminations considered for $ZnAl_2O_4$ are shown in Supplementary Figs. SI-1-3. Fig. 1 shows the typical termination of the (100), (110), and (111) surface slabs. However, deviations in axial ratios and interaxial angles from the cubic lattice forces the proposed algorithm to make models with different terminations in some surfaces, which are given in Fig. SI-4. Moreover, a reasonable termination for the (110) surface of $ZnMn_2O_4$ was not generated from the algorithm. The two-fold coordinated Zn at the (100) surface was additionally tilted toward the surface to lower the surface energy except for $ZnMn_2O_4$ and $ZnFe_2O_4$ where tilting did not lower the surface energy. Table 2 shows the calculated $E_{surf}$, difference between the number of spin up and down electrons (spin) per supercell, ionization potential (IP), and electron affinity (EA) for slabs with two different thicknesses. The bulk-based definition in Ref. [58] was used to obtain IPs and EAs.

Table 2. Calculated $E_{surf}$, difference between the number of spin up and down electrons (spin), IP, and EA. Units of spin is electron magnetic moment per cell (two surfaces).

| Surface (cubic) | Compound | $E_{surf}$ (meV/Å²) | | | Spin | | IP(eV) | | EA (eV) | |
|---|---|---|---|---|---|---|---|---|---|---|
| | | Fit | Thin | Thick | Thin | Thick | Thin | Thick | Thin | Thick |
| (100) | ZnAl$_2$O$_4$ | 102.8 | 94.9 | 93.8 | 0.0 | 0.0 | 6.6 | 6.6 | 2.2 | 2.2 |
| | ZnGa$_2$O$_4$ | 85.5 | 79.2 | 78.3 | 0.0 | 0.0 | 6.9 | 6.9 | 4.2 | 4.2 |
| | ZnIn$_2$O$_4$ | 70.4 | 65.6 | 65.0 | 0.0 | 0.0 | 6.5 | 6.5 | 5.2 | 5.1 |
| | ZnV$_2$O$_4$ | 83.7 | 77.9 | 77.1 | 0.1 | -0.1 | 4.3 | 4.4 | 3.7 | 3.8 |
| | ZnCr$_2$O$_4$ | 112.6 | 99.5 | 97.7 | 0.0 | 0.0 | 5.5 | 5.5 | 3.1 | 3.1 |
| | ZnMn$_2$O$_4$ | 56.9 | 46.4 | 44.9 | 0.1 | 0.0 | 5.5 | 5.4 | 4.7 | 4.6 |
| | ZnFe$_2$O$_4$ | 70.9 | 64.9 | 64.0 | 0.0 | 0.0 | 5.6 | 5.6 | 4.5 | 4.5 |
| | ZnCo$_2$O$_4$ | 102.6 | 97.6 | 96.8 | 0.1 | 0.2 | 5.6 | 5.6 | 3.7 | 3.7 |
| (110) | ZnAl$_2$O$_4$ | 134.5 | 126.4 | 124.8 | 0.0 | 0.0 | 6.9 | 6.9 | 2.5 | 2.5 |
| | ZnGa$_2$O$_4$ | 109.6 | 103.0 | 101.6 | 0.0 | 0.0 | 7.0 | 7.0 | 4.3 | 4.3 |
| | ZnIn$_2$O$_4$ | 91.9 | 86.7 | 85.7 | 0.0 | 0.0 | 6.5 | 6.5 | 5.1 | 5.1 |
| | ZnV$_2$O$_4$ | 94.3 | 103.3 | 105.1 | 0.7 | 0.0 | 4.5 | 4.4 | 3.9 | 3.9 |
| | ZnCr$_2$O$_4$ | 132.9 | 126.6 | 125.3 | 1.5 | 1.5 | 5.6 | 5.6 | 3.2 | 3.2 |
| | ZnFe$_2$O$_4$ | 86.6 | 80.8 | 79.6 | 0.0 | 0.0 | 5.7 | 5.7 | 4.5 | 4.5 |
| | ZnCo$_2$O$_4$ | 120.5 | 115.7 | 114.8 | 0.0 | 0.0 | 5.8 | 5.8 | 3.9 | 3.9 |
| (111) | ZnAl$_2$O$_4$ | 160.6 | 150.3 | 147.7 | 0.0 | 0.0 | 5.0 | 5.0 | 0.6 | 0.6 |
| | ZnGa$_2$O$_4$ | 128.2 | 120.2 | 118.2 | 1.3 | 0.6 | 5.9 | 5.9 | 3.1 | 3.1 |
| | ZnIn$_2$O$_4$ | 94.1 | 88.3 | 86.8 | 0.6 | 0.6 | 6.1 | 6.1 | 4.7 | 4.7 |
| | ZnV$_2$O$_4$ | 95.4 | 84.2 | 81.4 | 4.0 | 4.0 | 3.6 | 3.6 | 3.0 | 3.0 |
| | ZnCr$_2$O$_4$ | 128.1 | 119.7 | 117.6 | 0.0 | 0.0 | 4.4 | 4.4 | 1.9 | 1.9 |
| | ZnMn$_2$O$_4$ | 78.2 | 70.8 | 68.9 | 4.0 | 4.0 | 4.1 | 4.1 | 3.3 | 3.3 |
| | ZnFe$_2$O$_4$ | 94.4 | 86.5 | 84.5 | 0.0 | 0.0 | 4.8 | 4.8 | 3.7 | 3.7 |
| | ZnCo$_2$O$_4$ | 120.9 | 112.9 | 111.0 | 4.0 | 4.0 | 4.6 | 4.6 | 2.7 | 2.7 |

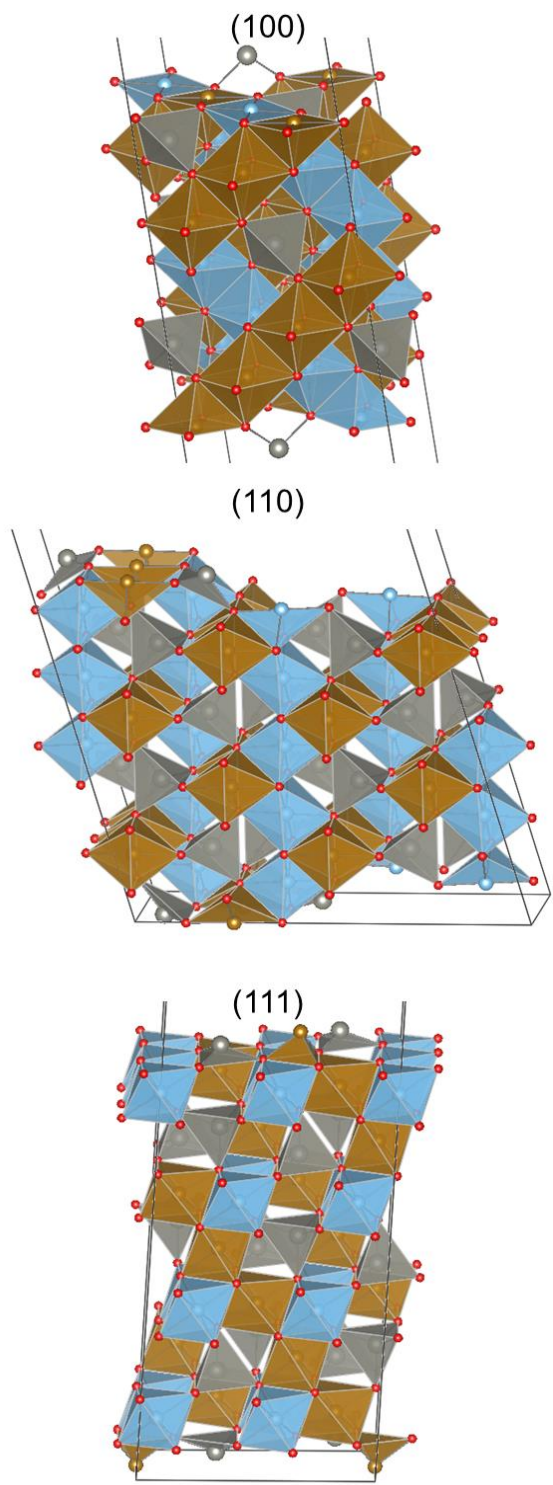

**Fig. 1.** Terminations of ZnM$_2$O$_4$ surfaces. Gray, brown, blue, and red balls indicate Zn, spin-up M, spin-down M, and O. Spin up and down are not distinguished for M=Al, Ga, In, and Co.

## 3.2 O vacancy formation

O desorption calculations were conducted with a double supercell of those used in surface calculations without defects. The spacing between defects in these models is anticipated to sufficiently converge defect formation energies based on our previous work[21]. $E_{Ovac}$ for various O sites are shown in Figs. SI-5-12. For the (100) surface, all M are in square pyramidal coordination with five O, and the surface O are at the base of two pyramids. Some surface O additionally bond to Zn, though there is no clear rule relating the value of $E_{Ovac}$ and existence of bonds to Zn from the vacancy site. There are three types of bonding environments in O at the (110) surface: (a) two-fold coordination with a four-fold coordinated M and a four-fold coordinated Zn, (b) three-fold coordination with a four-fold coordinated M, a six-fold coordinated M, and a three-fold coordinated Zn, and (c) three-fold coordination with three six-fold coordinated M. On the other hand, there are three types of bonding environments in O at the (111) surface: (a) three-fold coordination with two six-fold coordinated M and a three-fold coordinated Zn, (b) three-fold coordination with three six-fold coordinated M, and (c) four-fold coordination with a three-fold coordinated M, two six-fold coordinated M, and a three-fold coordinated Zn.

### 3.2.1 Discussion on smallest $E_{Ovac}$ for each surface

First we discuss the smallest $E_{Ovac}$ for each surface (smallest $E_{Ovac}$) because the minimum $E_{Ovac}$ site is where reactions are most likely to occur. Figure 2 plots the lowest $E_{Ovac}$ for each surface versus various quantities that can be obtained without explicit defect calculations. Figure 2a shows the smallest $E_{Ovac}$ versus $E_{surf}$ where points for the same M are connected with lines. The difference in $E_{Ovac}$ between the most stable and least stable surfaces span as much as 3.6 eV in $ZnGa_2O_4$, strongly indicating that the surface stability must be specified when discussing the smallest $E_{Ovac}$ of the same material. $E_{surf}$ is smallest for the (100) surface (red point is leftmost) for all M, while the smallest $E_{Ovac}$ is largest in the (100) surface (red point is highest) for all compounds except for $ZnCo_2O_4$. Connected points for the same M goes down toward the right side with the exception of $ZnCo_2O_4$, indicating that the smallest $E_{Ovac}$ of a surface of a material tends to be larger if the surface energy is lower. This finding is reasonable because less stable surfaces would have sites that are in more awkward environments and therefore vulnerable to removal. The points align along a concave curve rather than a straight line. Figure 2b shows the smallest $E_{Ovac}$ versus BG. A decent correlation with the smallest $E_{Ovac}$ and BG is visible in the (100) and (110) surfaces of zinc spinels though $ZnV_2O_4$ is an outlier. A close observation of the electronic structure of $ZnV_2O_4$ reveals that there are substantial contributions from O at the bottom of the conduction band in the investigated zinc spinels except for $ZnV_2O_4$ where the conduction band minimum (CBM) is V $3d$ ($t_{2g}$) states at 0.6 eV from the valence band maximum (VBM) (Figs. SI-13-14). An assumption is made that two bands between 0 eV and 2 eV above the VBM are purely V majority spin $3d$ ($t_{2g}$) bands and the lowest O states in the conduction band overlaps with states in the V minority spin $3d$ ($t_{2g}$) band about 2.1 eV from the VBM. Using a correction of 1.5 eV in the CBM of $ZnV_2O_4$, which effectively sets the CBM at the V minority spin $3d$ band minimum, the BG becomes 2.1 eV and the smallest $E_{Ovac}$ of 4.2 eV/defect in $ZnV_2O_4$ becomes in line with the trend formed by other zinc spinels (this correction is applied in empty symbols in Figs. 2b,d,f). $E_{Ovac}$ for the (111) $ZnAl_2O_4$ and $ZnGa_2O_4$ surfaces appear to be outliers that are too low in Fig. 2b-d, but the reason is not clear. Plots of the smallest $E_{Ovac}$ versus $E_{bulk}$, EA, IP, and the work function (WF), which is the mean average of the IP and EA, are given in Figs. 2c-f, respectively.

The correlation between smallest $E_{Ovac}$ of a surface and other quantities can be quantified using the coefficient of determination ($R^2$) as shown in Table 3. The (100) and (110) surfaces showed a decent correlation ($R^2 > 0.5$) between the smallest $E_{Ovac}$ and $E_{bulk}$ as well as EA. $R^2 < 0.5$ for BG, but the corrected BG when the CBM of $ZnV_2O_4$ was taken at the V minority spin $3d$ band minimum (BG') satisfies $R^2 > 0.5$. Here, $E_{bulk}$, EA, and BG are the three quantities with $R^2 > 0.5$ in our previous study on binary oxides [21], and a good correlation between the surface anion vacancy and $E_{bulk}$ was found in group 3 to 5 hydrides, carbides, and nitrides [59]. On a side note, the bulk O vacancy formation energy of oxides could be modeled by a linear combination of the enthalpy of formation, O $2p$ band center, band gap, and electron negativity difference between cation and anion[60]. In contrast, only the IP and WF have $R^2 > 0.5$ in the (111) surface where the smallest $E_{Ovac}$ of $ZnAl_2O_4$ and $ZnGa_2O_4$ is very low compared to what is expected from the trend with $E_{bulk}$, EA, and BG from other M. The largest $R^2$ (0.86) in Table 3 is between the smallest $E_{Ovac}$ and IP of the IP (111) surface, which is difficult to explain based on chemical intuition.

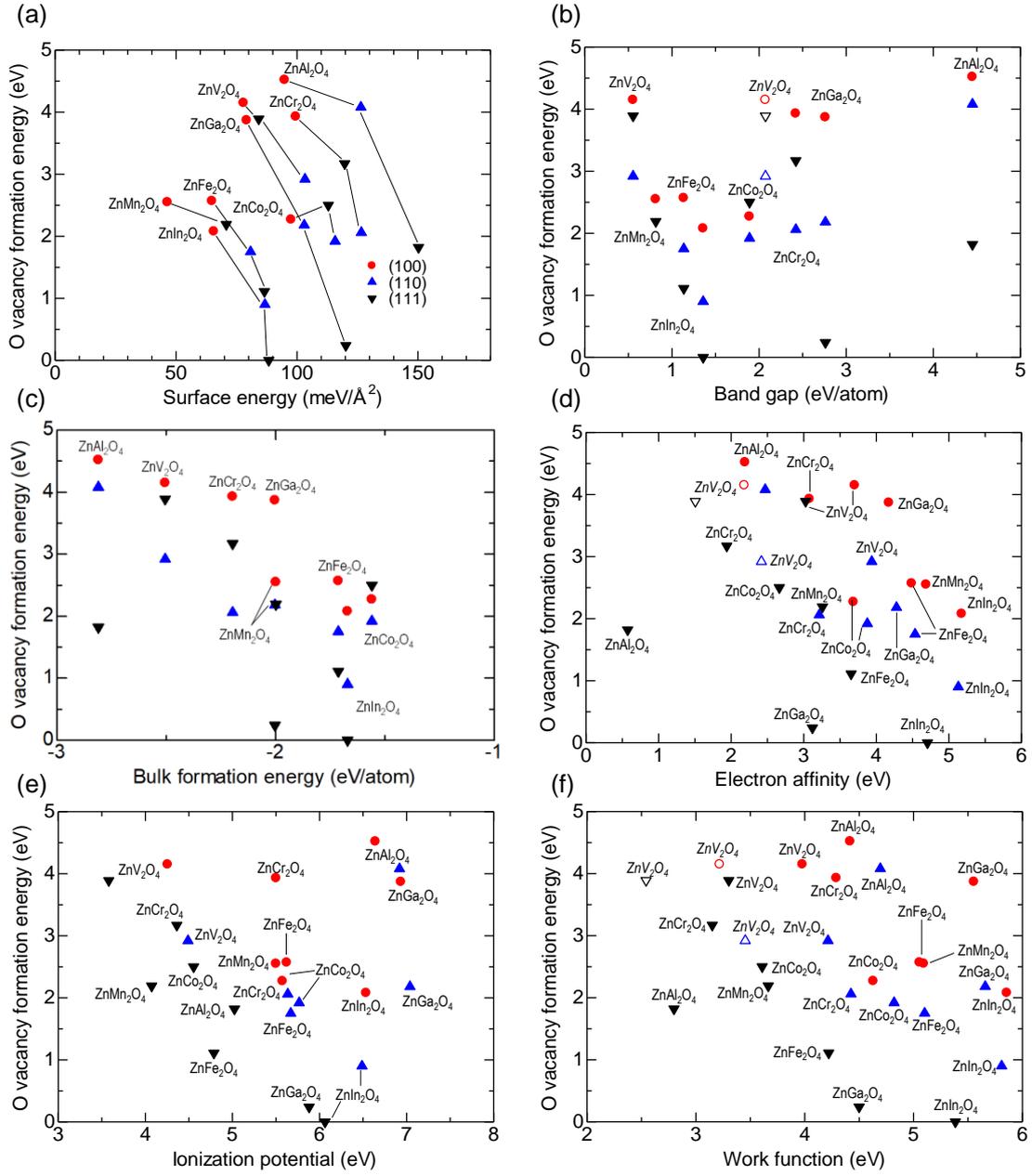

Fig. 2. (a) $E_{surf}$, (b) minimum BG, (c) $E_{bulk}$, (d) EA, (e) IP, and (f) WF versus the smallest $E_{Ovac}$ of a surface. The empty symbols in (b,d,f) is when the V minority spin $3d$ ($t_{2g}$) band bottom is taken as the CBM in $ZnV_2O_4$.

Table 3. Coefficient of determination ($R^2$) of the smallest $E_{Ovac}$ versus other quantities. Primed values are when the V minority spin $3d$ ($t_{2g}$) band bottom is taken as the CBM in $ZnV_2O_4$.

| Surface | BG | BG' | $E_{surf}$ | $E_{bulk}$ | EA | EA' | IP | WF | WF' |
|---|---|---|---|---|---|---|---|---|---|
| (100) | 0.31 | 0.64 | 0.25 | 0.81 | 0.59 | 0.69 | 0.00 | 0.34 | 0.34 |
| (110) | 0.37 | 0.74 | 0.38 | 0.80 | 0.72 | 0.76 | 0.00 | 0.34 | 0.30 |
| (111) | 0.04 | 0.00 | 0.00 | 0.19 | 0.20 | 0.44 | 0.86 | 0.64 | 0.78 |

3.2.1 Discussion on all $E_{Ovac}$ for each surface and element M

We turn to discussion on relations regarding all O vacancy sites. Figure 3 shows the $E_{Ovac}$ for all surface O sites and all $ZnM_2O_4$ plotted against $E_{bulk}$ of $ZnM_2O_4$. Different symbols are used for different orientations. The $R^2$ for all points is 0.36, and a higher $E_{Bulk}$ tends to result in a higher $E_{Ovac}$, which is a trend also found in binary oxides.[21] The correlation between $E_{Ovac}$ and $E_{bulk}$ strongly depends on the orientation; $R^2$ is highest in (100) at 0.78, relatively high in (111) at 0.54, and lowest in (111) at 0.18. $E_{bulk}$ had the highest feature importance in (100) and (110) but not in (111) for smallest $E_{Ovac}$ in each $ZnM_2O_4$, which is the same trend as the $R^2$ for all $E_{Ovac}$.

There is an obvious site environment dependence on $E_{Ovac}$ in (110) and (111). As shown in Figs. SI-5-12, two-fold coordinated O sites at the (110) surface have lower $E_{Ovac}$ than three-fold coordinated sites. The absolute value of the difference between the average $E_{Ovac}$ of three-fold and two-folds coordinated O sites lie between 0.86 to 1.88 eV (in In and Ga, respectively). The percentage of the difference of two-fold sites against three-fold sites is between 28% to 48% (in Al and Ga, respectively). Similarly, the average $E_{Ovac}$ of four-fold coordinated O sites at the (111) surface is 0.06 to 1.67 eV (in V and Ga, respectively) higher than the average of three-fold coordinated O sites. The range of the percentage is 1 to 356% (V and Ga, respectively). The geometrical contribution to the $E_{Ovac}$ of the (111) surface, namely the effect of coordination number, therefore differs widely over the species of M in $ZnM_2O_4$ in (111), resulting in the lower correlation between $E_{Ovac}$ and $E_{bulk}$. The correlation between $E_{Ovac}$ and $E_{bulk}$ ignores all site-dependent contributions, thus large scattering between species of M in the geometrical contribution should result in a low correlation. In contrast. the ratio of $E_{Ovac}$ between two-fold and three-fold coordinated O sites on the (110) surface is mostly the same between species of M, thus the site-dependent contributions mostly cancel out and result in a much larger $R^2$ between $E_{Ovac}$ and $E_{bulk}$ when compared to the (111) surface (Fig. 3). There is a trend in the relation between the position of M in the periodic table and $E_{Ovac}$; $E_{bulk}$ increases (absolute value decreases) and $E_{Ovac}$ tends to decrease as the row number increases in group 13 elements and as the group number increases in 3$d$ elements.

The relations between site descriptors and the $E_{Ovac}$ in ZnAl$_2$O$_4$ and ZnCr$_2$O$_4$ are shown in Fig. 4. The absolute value of the Bader charge, as obtained by the bader code[61-64], the average bond length, and the coordination number of the desorbing O site were considered as site descriptors. Although a comparison of Bader charge against $E_{Ovac}$ for all O sites over all orientation and all ZnM$_2$O$_4$ is very tempting, this is actually a very bad idea. When pseudopotentials are used, as in this study, the Bader volume strongly depends on the number of valence electrons in each element. In an extreme case, using $n$ valence electrons for an atom with a nominal valence of +$n$ results in a nominal charge density of zero for the atom. In such a case, the Bader volume could be zero. However, adding additional valence electrons results in a peak in charge density near the center of the atom, resulting in a non-zero Bader volume. Using the charge density of all electrons is another option, but the total number of electrons based on charge integration tend to differ from the intended number. The reason is a drastic change in the charge density near the nuclei that is difficult to integrate for heavy elements; avoiding this problem is a strong motivation to use pseudopotentials. The choice of pseudopotential is expected to cancel out for the same compound. In contrast, the average bond length and O coordination number are transferrable descriptors that can be used for any system. The results for Bader charge, average bond length, and O coordination number are shown in Figs. SI-15-17, respectively. The average O-Zn bond length could not be used as a descriptor because there are no O-Zn bonds in some O sites. Needless to say, $E_{bulk}$ and other bulk descriptors are the same for all O sites in the same ZnM$_2$O$_4$, therefore cannot be used. The goodness of correlation between $E_{Ovac}$ and site descriptor differed significantly over different ZnM$_2$O$_4$, and Figs. 4(a,c,e) show plots of ZnAl$_2$O$_4$ with very bad correlation and Figs. 4(b,d,f) are plots of ZnCr$_2$O$_4$ with good correlation.

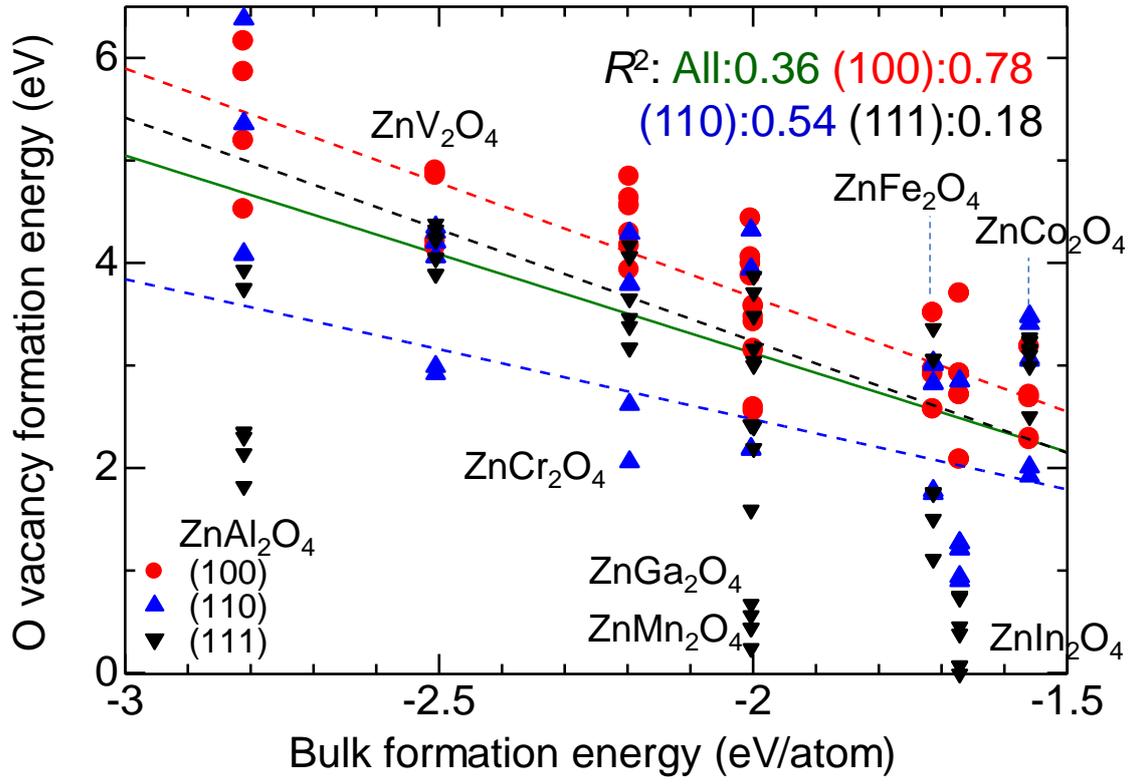

Fig. 3. $E_{Ovac}$ for all surface O sites and all ZnM$_2$O$_4$ plotted against $E_{bulk}$ of ZnM$_2$O$_4$.

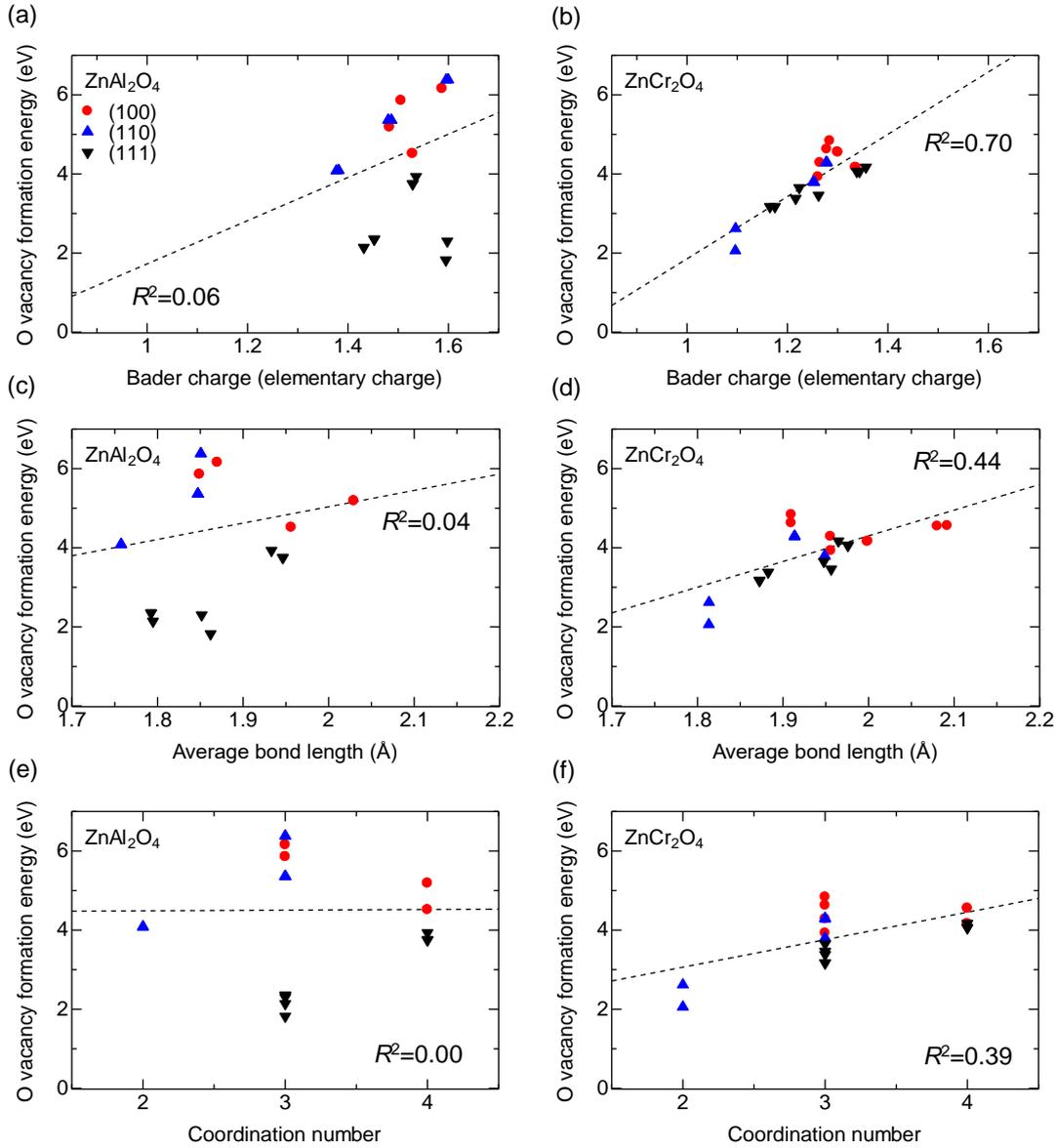

Fig. 4. $E_{Ovac}$ versus (a,b) O site Bader charge, (c,d) average bond length, and (e,f) coordination number of desorbing O for (a,c,e) $ZnAl_2O_4$ and (b,d,f) $ZnCr_2O_4$.

In all $ZnM_2O_4$, the trend was a smaller $E_{Ovac}$ when the Bader charge was small (closer to zero), bond length was small, and coordination number was small. Removing an O anion as neutral species forces excitation of atoms in the valence band to a defect state above the VBM. A small Bader charge is therefore qualitatively consistent with a smaller $E_{Ovac}$ because a smaller amount of charge is excited, requiring less energy than when more charge must be excited. The coordination number is expected to correlate with $E_{Ovac}$ because desorption after severing fewer bonds should result in a lower desorption energy than sites with more bonds, which are sites with higher coordination number. The low correlation of the coordination number and $E_{Ovac}$ could be attributed to the small number of coordination number choices (two for every orientation) and the existence of additional factors that determine $E_{Ovac}$. We could not suggest a reason for a direct link between bond length and $E_{Ovac}$. However, the bond length and coordination number have a high $R^2$ of between 0.60 in $ZnCo_2O_4$ and 0.84 in $ZnFe_2O_4$, where a lower coordination number tends to reduce the average bond length. This is reasonable because a lower coordination number is expected to strengthen each bond and therefore shorten bond lengths. The bond length is thereby effectively acting as a descriptor of the coordination environment. Shorter bond lengths represent a smaller coordination number that need to be severed and, therefore, smaller $E_{Ovac}$.

Use of further geometrical descriptors may appear interesting but are difficult to implement. An ideal geometrical descriptor reflects the entire coordination environment in some form rather than a limited aspect of the coordination, such as minimum, maximum, difference between maximum-minimum, or average bond length or bond angle. The distortion of the coordination was identified as a good geometrical descriptor of $E_{Ovac}$ in three-fold coordinated O in θ-$Al_2O_3$ and β-$Ga_2O_3$. This distortion is a single quantity that incorporates information of all bond lengths and angles and therefore reflects the three-dimensional coordination environment. The basic idea is that a large distortion of the coordination environment from an ideal $sp^2$ or $sp^3$ coordination makes the O site relatively unstable and therefore decreases $E_{Ovac}$.[65] However, O in $ZnM_2O_4$ bonds with tetrahedrally coordinated Zn and octahedrally coordinated M, thus the coordination environment is much more complicated. As a result, we compromised with the average bond length, which actually did have some correlation with $E_{Ovac}$, as a descriptor containing information on all bond lengths.

Finally, statistical analysis based on machine learning (ML) techniques were also carried out to predict $E_{Ovac}$ for all the surface O sites of $ZnM_2O_4$ and identify the important factors for their prediction. Descriptors discussed above such as $E_{bulk}$, IP, EA, BG, and geometrical descriptors were used. Types of the surface orientations were also implemented using a one-hot encoding method, one example is the Surface(111) descriptor, which is 1 for an O vacancy on the (111) surface and 0 otherwise. Evaluations of well-performing ML methods were performed with a set of six widely used ML methods including three major categories: linear methods for linear regression, kernel methods and tree ensemble methods for nonlinear regression. More specifically, we tested least absolute shrinkage and selection operator (LASSO) and ordinary linear squares (OLS) regressions as linear methods, support vector regression (SVR) and Gaussian process regression (GPR) as kernel methods, and random forest regression (RFR) and extra trees regression (ETR) as tree ensemble methods. To evaluate the predictive capability of the ML models, Monte Carlo cross validation with 100 times of random leave-10%-out trials was performed to obtain the average root-mean-square error (RMSE). Figure 5 demonstrates that the six ML methods tested in this study could predict the $E_{Ovac}$ within RMSE of 0.49-0.77 eV/defect. Tree ensemble methods predicted relatively well and ETR gave the best predictive accuracy among the models tested. The $R^2$ value was also calculated to be 0.82 for this ML model based on ETR. This result demonstrates that $E_{Ovac}$ can be predicted by using only 106 datapoints as a dataset and readily available descriptors. The accuracy can be improved once more data are calculated in the future.

With the best ML method (ETR) in hand, the Shapley additive explanations (SHAP) method (version 0.37.0) [66-68] was used to identify and prioritize descriptors, as shown in Fig. 6. Namely, contribution of a given input feature to the target ($E_{Ovac}$) response was identified. The most important descriptor was $E_{bulk}$, followed by the type of the surface orientation (Surface(111)), IP, coordination numbers and bond length average. As expected, $E_{bulk}$ is a very important descriptor, and the Surface(111) descriptor has a high contribution because, unlike the other two surfaces, the (111) surface has low correlation between $E_{Ovac}$ and $E_{bulk}$ (see Fig. 3 and Table 3). The IP has a good correlation with the smallest $E_{Ovac}$ in the (111) surface, but not in (100) and (111) (see Table 3). The analysis also revealed that $E_{Ovac}$ (SHAP value) tends to be high when $E_{bulk}$ (feature value) is low. This result indicates that information on not only bulk properties but also local structures of the surface O sites is necessary. In addition, both electronic and geometrical properties were found to be important for predicting $E_{Ovac}$. As an added note, Surface(111) is a discrete, one hot descriptor, thus points are shown in one of two colors (red and blue) in Fig. 6. However, the SHAP value is the contribution to the output, and therefore is not discrete

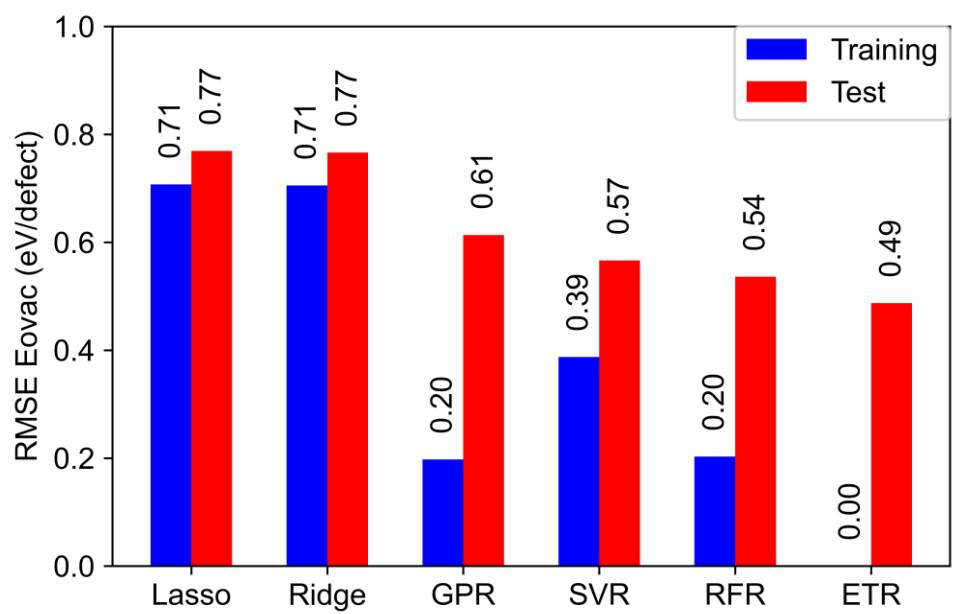

Fig. 5. Average RMSEs for predicting $E_{Ovac}$ for all the surface O sites of $ZnM_2O_4$ by 100 times of random leave-10%-out trials with various ML methods.

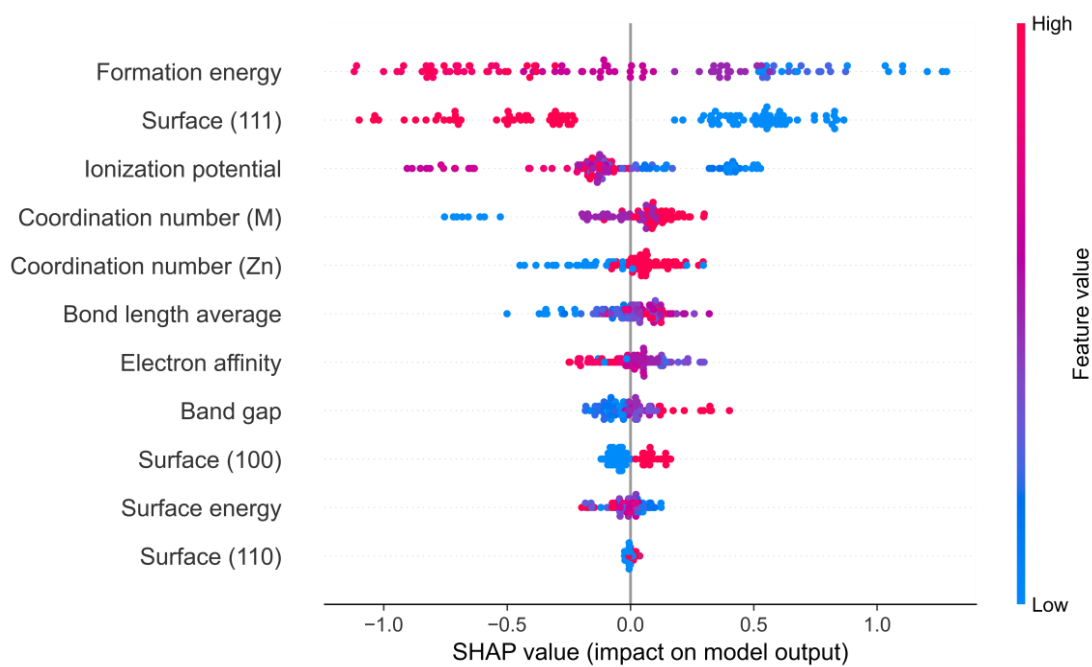

Fig. 6. SHAP values of the descriptors in predicting $E_{Ovac}$ using ETR. SHAP values for individual factors are plotted as dots (blue corresponds to low features, red to high features). Here, features are ordered in descending order according to the sum of the absolute values of the SHAP values.

In summary, a general trend of higher $E_{Ovac}$ with smaller $E_{bulk}$ was observed regardless of the orientation and O desorption site. The quantity $E_{bulk}$ could be a good descriptor of the surface anion desorption energy in not only oxides but in other compounds because $E_{bulk}$ is related to the bond strength between cations and anions and severing of such bonds is necessary to remove surface anions[21,59]. The Bader charge and the average bond length of an O site could act as a good descriptor of $E_{Ovac}$, regardless of orientation, in some systems. However, the $E_{Ovac}$ may be strongly off-trend in specific coordination environments that need to be evaluated case-by-case.

## 4. Conclusions

The $E_{Ovac}$ for various surface orientations of eight normal zinc-based spinel oxides $ZnM_2O_4$ (M is one of Al, Ga, In, V, Cr, Mn, Fe, or Co) were systematically evaluated and the correlation between physical quantities such as $E_{bulk}$, BG, and EA were investigated. A large variation of up to 3.6 eV in the smallest $E_{Ovac}$ of a surface was observed for different orientations of same material. $E_{Ovac}$ was typically higher in a more stable surface within the same compound, which is in line with chemical intuition. A good correlation between $E_{Ovac}$ and $E_{bulk}$, BG, and EA was obtained between the same orientation for (100) and (110) surfaces, although the trend for the (111) surface was contradictory and the IP was more important. These characteristics were reflected in machine learning of $E_{Ovac}$ of all surface sites in all orientations and all compounds. The $E_{bulk}$, a Boolean descriptor on whether the surface is (111), and the IP were identified as important descriptors, followed by geometrical descriptors that are different in each O site.

**Acknowledgments**

This study was funded by a grant (No. JPMJCR17J3) from CREST of the Japan Science and Technology Agency (JST), and by a Kakenhi Grant-in-Aid (No. 18K04692) from the Japan Society for the Promotion of Science (JSPS). Computing resources of the Research Institute for Information Technology at Kyushu University, ACCMS at Kyoto University, and the Supercomputer Center in the Institute for Solid State Physics at the University of Tokyo were used. The VESTA code[69] was used to draw Fig. 1 and Supplementary Figs. SI-1-12.